\pdfoutput=1
\RequirePackage{ifpdf}
\ifpdf 
\documentclass[pdftex]{sigma}
\else
\documentclass{sigma}
\fi

\numberwithin{equation}{section}

\begin{document}
\allowdisplaybreaks

\newcommand{\arXivNumber}{1912.07952}

\renewcommand{\PaperNumber}{034}

\FirstPageHeading

\ShortArticleName{Breathing Modes, Quartic Nonlinearities and Effective Resonant Systems}

\ArticleName{Breathing Modes, Quartic Nonlinearities\\ and Effective Resonant Systems}

\Author{Oleg EVNIN~$^{\dag\ddag}$}

\AuthorNameForHeading{O.~Evnin}

\Address{$^\dag$~Department of Physics, Faculty of Science, Chulalongkorn University, Bangkok, Thailand}

\Address{$^\ddag$~Theoretische Natuurkunde, Vrije Universiteit Brussel and the International Solvay Institutes, \\
\hphantom{$^\ddag$}~Brussels, Belgium}
\EmailD{\href{mailto:oleg.evnin@gmail.com}{oleg.evnin@gmail.com}}

\ArticleDates{Received February 20, 2020, in final form April 14, 2020; Published online April 23, 2020}

\Abstract{A breathing mode in a Hamiltonian system is a function on the phase space whose evolution is exactly periodic for all solutions of the equations of motion. Such breathing modes are familiar from nonlinear dynamics in harmonic traps or anti-de Sitter spacetimes, with applications to the physics of cold atomic gases, general relativity and high-energy physics. We discuss the implications of breathing modes in weakly nonlinear regimes, assuming that both the Hamiltonian and the breathing mode are linear functions of a coupling parameter, taken to be small. For a linear system, breathing modes dictate resonant relations between the normal frequencies. These resonant relations imply that arbitrarily small nonlinearities may produce large effects over long times. The leading effects of the nonlinearities in this regime are captured by the corresponding effective resonant system. The breathing mode of the original system translates into an exactly conserved quantity of this effective resonant system under simple assumptions that we explicitly specify. If the nonlinearity in the Hamiltonian is quartic in the canonical variables, as is common in many physically motivated cases, further consequences result from the presence of the breathing modes, and some nontrivial explicit solutions of the effective resonant system can be constructed. This structure explains in a uniform fashion a series of results in the recent literature where this type of dynamics is realized in specific Hamiltonian systems, and predicts other situations of interest where it should emerge.}

\Keywords{weak nonlinearity; multiscale dynamics; time-periodic energy transfer}

\Classification{35B20; 35Q55; 35Q75; 35L05; 81Q05}

\section{Introduction}

Even in a complicated nonlinear dynamical system, with chaotic trajectories and all, it may turn out that some specific combinations of the dynamical variables always behave periodically with the same period, independent of the initial conditions. In such situations, one is talking about {\it breathing modes}.\footnote{One should distinguish breathing modes from cases where trajectories themselves are periodic, associated with superintegrability, such as the periodic trajectories of the Kepler problem. None of the systems that motivate our study are known to be superintegrable, nor even integrable. Their trajectories may be arbitrarily complicated, and are certainly not exactly periodic, while a breathing mode only gives one specific function of the phase space variables that is periodic for all trajectories.} Perhaps the simplest example is the separation of the center-of-mass motion in a harmonic potential~\cite{bbbb}, say, for a system of identical classical particles with arbitrary translationally invariant 2-body interactions. The center-of-mass always behaves like a single independent particle bound by a harmonic potential, and all of its trajectories are periodic with the same period, providing a breathing mode. There are less obvious examples, such as the Pitaevskii--Rosch breathing mode \cite{breathing,qbreathing} for two-dimensional Bose--Einstein condensates in a~harmonic potential, as well as relativistic analogs of these systems involving nonlinear wave equations in anti-de Sitter spacetimes that we shall comment upon below. Recent experimental work on cold atomic gases motivated by breathing modes can be found in \cite{tri}.

The most immediate consequence of the breathing modes is in existence of time-dependent symmetry transformations in the system (kinematic symmetries) akin to Galilean and Lorentz boosts, except that the dependence on time is periodic rather than linear. For systems in harmonic traps, these symmetries have been explored, for instance, in \cite{Niederer, OFN}. Such symmetries are generated by the breathing modes in the same manner as ordinary time-independent symmetry transformations are generated by conserved quantities. One can of course use these symmetries to construct new solutions, as has been done in \cite{translations} to produce time-dependent solutions from stationary configurations. It may seem that there is nothing more profound in this than boosting Galilean-invariant systems from one inertial frame to another, though in practice the effect of the transformations generated by the known breathing modes may be much less evident than in the Galilean case. Our focus in this article, however, will be precisely on the implications of breathing modes for the dynamics that go beyond mere applications of `boosts.'

As with any symmetries, the presence of breathing modes imposes strong constraints on the Hamiltonian of the system. For a linear system, it mandates resonant relations between the frequencies of the normal modes, which form evenly spaced ladders. If weak nonlinearities are turned on, the resonances between the linearized normal modes result in an enhancement of nonlinear interactions, so that a nonlinearity of order $g\ll 1$ may induce arbitrarily strong effects on long timescales of order $1/g$. A standard way to accurately capture the leading effects in this regime is the resonant approximation, also known as the multiscale analysis, effective equation or time-averaging method \cite{KM, murdock}, which simply discards all nonresonant mode couplings, irrelevant at small~$g$. Our key result is that, under a simple condition, breathing modes result in ordinary conserved quantities within the resonant approximation. We then focus on a situation generic for interacting field theories, relativistic and non-relativistic, where the leading nonlinearity is quartic in the dynamical variables, that is, a Hamiltonian $H=H_0+gH_1$ that admits a~breathing mode $B=B_0+g B_1$ so that $H_0$ and $B_0$ are quadratic in the dynamical variables and~$H_1$ is quartic. Making a simple assumption about the Poisson brackets of~$B_0$ and its complex conjugate, which essentially means that the symmetry algebra closes without generating extra conserved quantities apart from the already known ones, we obtain strong constraints on the resonant system corresponding to~$H$ at small~$g$. In this formulation, $B_0$ becomes an exact conserved quantity of the resonant system, while a family of explicit analytic solutions can be constructed within the resonant system at the full nonlinear level, accurately approximating solutions of the original system on long time scales of order $1/g$. This is an example where the presence of breathing modes constrains the system and allows one to construct novel analytic solutions that do not follow from applying the symmetry transformations generated by the breathing mode to any obvious solutions.

The structures outlined above have been observed in the recent literature for a number of special cases, motivated by rather disparate topics in physics. Thus, the analysis of \cite{BBCE,BBCE1,GGT, GHT} is rooted in the Gross--Pitaevskii equation and the physics of Bose--Einstein condensates, while the analysis of \cite{CF,BEF,BHP,BHP1,BEL} originates in studies of nonlinear dynamics in anti-de Sitter spacetimes, which is of interest for mathematical general relativity and high-energy physics. (Similarities between these two classes of systems have been pointed out in~\cite{BMP} and explained via taking nonrelativistic limits in \cite{BEF, BEL}.) In the case of Bose--Einstein condensates, the existence of breathing modes is well-known \cite{bbbb,breathing,qbreathing}, though their relation to the weakly nonlinear solutions within the resonant approximation has not been duly appreciated in the literature. For the case of \cite{CF}, the solutions of the resonant approximation came first, and our present analysis will supply the corresponding breathing mode responsible for these solutions. The general type of resonant systems relevant for us here, where the presence of an extra conserved quantity imposes relations between mode couplings and generates some explicit analytic solutions, has been constructed in \cite{AO,AO1}. Our present exposition explains the origin of these structures in the underlying Hamiltonian dynamics from which the resonant approximation originates. Breathing modes are common in system with dynamical symmetry, where the Hamiltonian is realized as a Cartan generator of the corresponding dynamical symmetry Lie algebra. Likewise, quartic nonlinearities that play a significant role in our treatment are generic leading nonlinearities under the assumption that odd order nonlinearities are prohibited by a reflection symmetry in the configuration space. The framework we present here gives a recipe to search for explicit weakly nonlinear solutions within this class of systems.

\section{Breathing modes}\label{sec_breathe}

Consider a system with the Hamiltonian $H(p,q)$ and the usual equations of motion
\begin{gather*}
\frac{{\rm d}p_i}{{\rm d}t}=-\frac{\partial H}{\partial q_i},\qquad \frac{{\rm d}q_i}{{\rm d}t}=\frac{\partial H}{\partial p_i}.
\end{gather*}
A function $B(p,q)$ on the phase space is called a {\it breathing mode} if
\begin{gather}
\frac{{\rm d}B}{{\rm d}t}=\{H,B\}\equiv\sum_k\left(\frac{\partial H}{\partial p_k}\frac{\partial B}{\partial q_k}-\frac{\partial H}{\partial q_k}\frac{\partial B}{\partial p_k}\right)={\rm i}B.\label{breathdef}
\end{gather}
This equation is evidently solved by
\begin{gather*}
B(t)={\rm e}^{{\rm i}t}B(0),
\end{gather*}
and hence $B(p(t),q(t))$ oscillates for all solutions of the equations of motion with the same period equal $2\pi$. Note that whenever $\{H,B\}$ is proportional to ${\rm i}B$, we can always set it equal to~${\rm i}B$, as in~(\ref{breathdef}), by rescaling~$H$, and this is the normalization of the Hamiltonian we shall assume below without loss of generality.

Existence of breathing modes of the form~(\ref{breathdef}) is a strong restriction on the system (that we intend to exploit), but at the same time the algebraic structure of~(\ref{breathdef}) is completely generic for systems with dynamical symmetries. Indeed, if $\{H,\cdot\}$ is a generator of a dynamical Lie group lying in the Cartan subalgebra and $\{B,\cdot\}$ is a generator corresponding to a positive root, one gets a relation of the sort~(\ref{breathdef}). In such situations, many breathing modes can be present on the same footing, corresponding to different generators of the dynamical symmetry group, as is indeed the case for the systems that motivate our current study \cite{BBCE,BBCE1,CF,BEF,BHP,BHP1,BEL,GGT,GHT}. Nonetheless, one can often construct consistent dynamical truncations of such systems to a~subset of degrees of freedom, either at the level of the full system or at the level of the resonant approximation in the weakly nonlinear regime, so that only one breathing mode is relevant in each truncation, which is again what happens in \cite{BBCE,BBCE1,CF,BEF,BHP,BHP1,BEL,GGT,GHT}. We shall therefore focus here on systematically exploring the consequences of having one breathing mode, while keeping in mind that in cases with many breathing modes some extra work may have to be done to make our results applicable.

The breathing mode generates a kinematic symmetry given by
\begin{gather}
q_i\to q_i+\eta \frac{\partial B}{\partial p_i}+\bar\eta \frac{\partial\bar B}{\partial p_i},\qquad p_i\to p_i-\eta\frac{\partial B}{\partial q_i}-\bar\eta\frac{\partial \bar B}{\partial q_i},\label{breathtransf}
\end{gather}
where $\eta$ is a complex-valued infinitesimal parameter, and bars denote complex conjugation, here and for the rest of our treatment. Unlike the case of ordinary symmetries (whose generators have vanishing Poisson brackets with the Hamiltonian), these transformations do not commute with the evolution, but rather induce very simple, predictable changes in the dynamical trajectory. Namely, if one applies~(\ref{breathtransf}) at $t=0$, the subsequent trajectory is transformed as
\begin{gather*}
q_i(t)\to q_i(t)+{\rm e}^{{\rm i}t}\eta \frac{\partial B}{\partial p_i}+{\rm e}^{-{\rm i}t}\bar\eta \frac{\partial\bar B}{\partial p_i},\qquad p_i(t)\to p_i(t)-{\rm e}^{{\rm i}t}\eta\frac{\partial B}{\partial q_i}-{\rm e}^{-{\rm i}t}\bar\eta\frac{\partial \bar B}{\partial q_i}.
\end{gather*}
For the simplest case of systems in harmonic potentials, such transformations are discussed in \cite{translations, Niederer, OFN}.

Our main focus in this article will be on weakly nonlinear systems with a breathing mode, so that
\begin{gather}
H=H_0+g H_1,\qquad B=B_0+gB_1,\label{HBgdef}
\end{gather}
with $g\ll 1$. We shall assume that $H_0$ is quadratic in the dynamical variables (which simply means that $g=0$ corresponds to a linear system), and so is $B_0$ (which means that the corresponding kinematic symmetry for this linear system at $g=0$ is linearly realized). Substituting these expressions in (\ref{breathdef}) and equating the coefficients of different powers of $g$, we obtain\footnote{There is some similarity between these expressions and the `integrable matrix theory' of \cite{Sh,Sh1,ShYu}, where Hamiltonians and symmetry generators depending linearly on a coupling parameter are considered for quantum-mechanical systems with finite-dimensional Hilbert spaces. Our classical phase space functions are naturally replaced by matrices, and commutators take the place of our Poisson brackets. The only substantial difference is that the right-hand sides of (\ref{01comm}) would be zero in the framework of \cite{Sh,Sh1,ShYu}, since one is dealing with conserved quantities rather than breathing modes.}
\begin{gather}
\{H_0,B_0\}={\rm i}B_0,\qquad \{H_0,B_1\}+\{H_1,B_0\}={\rm i}B_1,\qquad \{H_1,B_1\}=0.\label{01comm}
\end{gather}
In the later parts of our analysis, we shall also be assuming that $H_1$ is quartic in the dynamical variables, which is a generic situation for classical field systems with a field-reflection symmetry (and corresponds to generic two-body interactions in the quantum case).

We conclude this section with a few examples of breathing modes in relativistic and non-relativistic field systems that fit the above framework:
\begin{itemize}\itemsep=0pt
\item Consider a classical complex nonrelativistic field in $D$ spatial dimensions with the Hamiltonian
\begin{gather}
H=\frac12\int {\rm d}^Dx\bigg[\partial_k\bar\Psi(x)\partial_k\Psi(x)+x^kx^k |\Psi|^2(x) \nonumber\\
\hphantom{H=}{} +g|\Psi|^2(x)\int {\rm d}^Dy \, V(x-y)|\Psi|^2(y)\bigg].\label{bosgen}
\end{gather}
The momenta conjugate to $\Psi(x)$ are understood to be ${\rm i}\bar\Psi(x)$, so that the Hamiltonian equations of motion take the form of a nonlinear Schr\"odinger equation. (The first two terms of the Hamiltonian may of course be equivalently rewritten in the vector notation as $|\nabla\Psi|^2+x^2|\Psi|^2$.)
Quantization of this Hamiltonian (which we do not consider here) would have led to a system of identical bosons in an external harmonic potential interacting via translationally invariant two-body interactions given by $V(x-y)$, which is a standard subject in the physics of cold atomic gases. The classical Hamiltonian given above describes, from this perspective, the regime in which the trapped bosons undergo Bose--Einstein condensation. There is a set of breathing modes associated to the center-of-mass motion in~$D$ spatial dimenstions:
\begin{gather}
B_n=\int {\rm d}^Dx \, \big(x_n|\Psi|^2- \bar\Psi\partial_n \Psi\big).\label{Bosc}
\end{gather}
Of particular importance in our context are combinations of these modes in the form $B_x+{\rm i}B_y$ (not necessarily in two dimensions) that play a role in the dynamics of the Landau level truncations \cite{BBCE,BBCE1,GGT} of the evolution corresponding to~(\ref{bosgen}).
\item In two spatial dimensions and for the case of contact interactions, the symmetries of (\ref{bosgen}) get enhanced \cite{OFN}. The corresponding Hamiltonian is
\begin{gather*}
H=\frac12\int {\rm d}x {\rm d}y \,\big(|\partial_x\Psi|^2+|\partial_y\Psi|^2+\big(x^2+y^2\big) |\Psi|^2 +g|\Psi|^4\big).
\end{gather*}
This is known to possess the Pitaevskii--Rosch breathing mode \cite{breathing,qbreathing}, which manifests itself in perfectly periodic evolution of
\begin{gather*}
I=\int {\rm d}x{\rm d}y \, \big(x^2+y^2\big) |\Psi|^2.
\end{gather*}
To recast this mode in our standard form (\ref{breathdef}), one introduces
\begin{gather}
B=(I-H)-\frac{1}{2}\int {\rm d}x{\rm d}y\,\big[ x\big(\bar\Psi\partial_x\Psi-\Psi\partial_x\bar\Psi\big)
+y\big(\bar\Psi\partial_y\Psi-\Psi\partial_y\bar\Psi\big)\big],\label{BPR}
\end{gather}
which satisfies ${\rm d}B/{\rm d}t=2{\rm i}B$. (This can be changed to ${\rm d}B/{\rm d}t={\rm i}B$ to literally match our definition~(\ref{breathdef}) by a simple rescaling of the Hamiltonian, as per our general discussion.)
\item We now formulate relativistic analogs of the above two cases. An analog of the harmonic potential is provided by anti-de Sitter (AdS) spacetimes (maximally symmetric spacetimes of constant negative curvature) that play the same role for relativistic wave equations as the harmonic potential does for nonlinear Schr\"odinger equations.

For \looseness=-1 $d$ spatial dimensions, we denote the corresponding AdS space as AdS$_{d+1}$. It can be realized as a hyperboloid in an auxiliary flat pseudo-Euclidean space of dimension $d+2$ parametrized by $\big(X,Y,X^k\big)$ with the line element ${\rm d}s^2= -{\rm d}X^2-{\rm d}Y^2+{\rm d}X^k{\rm d}X^k$, defined by
\begin{gather}
 -X^2-Y^2+X^kX^k=-1.\label{embedding}
\end{gather}
One can parametrize this embedded hyperboloid by $X^k$ and $t$ so that the two remaining embedding coordinates are given by
\begin{gather*}
X=\sqrt{1+X^kX^k}\cos t,\qquad Y=\sqrt{1+X^kX^k}\sin t.
\end{gather*}
The AdS metric can be the extracted as \cite{EN}
\begin{gather*}
{\rm d}s^2=-\big(1+X^kX^k\big){\rm d}t^2+\left(\delta_{ij}-\frac{X^iX^j}{1+X^kX^k}\right){\rm d}X^i{\rm d}X^j.
\end{gather*}
(Note that $t$ runs from 0 to $2\pi$ in the embedding~(\ref{embedding}), but as the AdS metric does not depend on~$t$, it can be straightforwardly extended to run from $-\infty$ to $\infty$, which is how the AdS space is normally understood.) One can now define a relativistic field theory in this space, which shares many properties of the nonlinear Schr\"odinger equation in a harmonic trap. We shall use a real scalar field $\phi({\bf X},t)$ and its conjugate momentum $\pi_\phi({\bf X},t)=\partial_t \phi/\big(1+X^kX^k\big)$, though a complex field could easily be employed if more contact with nonrelativistic theories is needed. The Hamiltonian is then $H=\int {\rm d}{\bf X} \, h({\bf X}; \pi_\phi,\phi)$ with
\begin{gather}
h({\bf X};\pi_\phi,\phi)=\frac12\left[\big(1+X^kX^k\big) \pi_\phi^2 + \partial_k\phi\partial_k\phi +\big(X^k\partial_k\phi\big)^2+m^2\phi^2+\frac{g}2\phi^4\right],\label{hAdS}
\end{gather}
as derived from the standard action $S=-\frac{1}2\int {\rm d}^{d+1}x \, \sqrt{-g}\big(g^{\mu\nu}\partial_\mu\phi\,\partial_\nu\phi+m^2\phi^2+g\phi^4/2\big)$.
Just like for a harmonic trap, the center-of-mass motion separates for any self-interactions respecting the AdS isometries (in particular, the $\phi^4$ interactions in the Hamiltonian above) and performs independent oscillations described by the breathing modes
\begin{gather}\label{breathAdS}
B_{n}=\int {\rm d}{\bf X} \left(\frac{X^n h}{\sqrt{1+X^kX^k}}+{\rm i}\sqrt{1+X^kX^k} \pi_\phi \partial_n\phi \right).
\end{gather}
\item Finally, there is a relativistic analog of the Pitaevskii--Rosch breathing mode that can be made manifest by considering the systems defined by~(\ref{hAdS}) in three spatial dimensions and with the value of~$m^2$ corresponding to a conformally coupled scalar~\cite{CF}. In fact, it is more convenient to consider the same type of scalar field on a spatial 3-sphere, which is related to the above AdS setup by a conformal transformation. We refer the reader to~\cite{CF} for detailed analysis of the corresponding equations. The important point for us here is that, restricting to scalar fields that only depend on time~$t$ and the polar angle~$x$ on the 3-sphere, and introducing $v(x,t)=\phi(x,t)\sin x$, one obtains the following nonlinear wave equation
\begin{gather*}
\partial_t^2 v-\partial_x^2 v+\frac{g v^3}{\sin^2x}=0
\end{gather*}
with the boundary conditions $v(0,t)=v(\pi,t)=0$. This equation possesses a breathing mode of the form
\begin{gather}
B = \int {\rm d}x \left[ \cos x\left((\partial_t v)^2+ (\partial_x v)^2 + \frac{g v^4}{2\sin^2x}\right) - 2 {\rm i}\sin x \partial_t v \partial_x v \right],
\label{BCF}
\end{gather}
which can, of course, equally well be expressed canonically through the momentum $\pi_v=\partial_t v$ conjugate to~$v$.
\end{itemize}

\section{Linear systems}

We start with setting $g=0$ in (\ref{HBgdef}) and considering a linear system with a quadratic Hamiltonian~$H_0$ and a quadratic breathing mode $B_0$. Any linear system performing bounded motion can be diagonalized and split into independent harmonic oscillators with normal frequencies $\omega_n>0$ described by the complex amplitudes $\alpha_n(t)={\rm e}^{{\rm i}\omega_n t}\alpha_n(0)$ whose canonically conjugate momenta are defined to be $-{\rm i}\bar\alpha_n(t)$. In these variables, any quadratic Hamiltonian corresponding to bounded motion becomes simply
\begin{gather}
H_0=\sum_n\omega_n\bar\alpha_n\alpha_n.\label{H0norm}
\end{gather}
We shall assume for the rest of our treatment that this diagonalization has been performed and our canonical variables are $\alpha_n$ and $-{\rm i}\bar\alpha_n$.

For this simple case, the structure of a general quadratic breathing mode $B_0$ can be made explicit.
Indeed, the most general possible expression is
\begin{gather*}
B_0=\sum_{nm}\big[b_{nm}\bar\alpha_n\alpha_m+b^{+}_{nm}\alpha_n\alpha_m+b^{-}_{nm} \bar\alpha_n\bar\alpha_m\big],
\end{gather*}
where $b_{nm}$, $b^+_{nm}$ and $b^-_{nm}$ are numbers. We have to impose
\begin{gather*}
\{H_0,B_0\}={\rm i}\sum_{k}\left(\frac{\partial H_0}{\partial\bar\alpha_k}\frac{\partial B_0}{\partial\alpha_k}-\frac{\partial H_0}{\partial\alpha_k} \frac{\partial B_0}{\partial\bar\alpha_k}\right)={\rm i}B_0,
\end{gather*}
which implies
\begin{gather*}
\omega_n b_{nm}-\omega_m b_{nm}=-b_{nm},\qquad (\omega_n+\omega_m)b^\pm_{nm}=\pm b^\pm_{nm}.
\end{gather*}
Since $\omega_n>0$, $b^-_{nm}=0$. The rest exclusively depends, at least at the level of linearized theory, on the spectrum of normal mode frequencies $\omega_n$. If there are two frequencies satisfying $\omega_n+\omega_m=1$, the corresponding $b^+_{nm}$ can have an arbitrary value. If there are two frequencies satisfying $\omega_m=\omega_n+1$, the corresponding~$b_{nm}$ can have an arbitrary value. (Evidently, $b_{nn}$ must be zero.)

While, within the linearized approximation, the above argument still leaves a huge amount of freedom in constructing breathing modes, provided that the spectrum $\omega_n$ satisfies simple constraints, it is worth discussing upfront which of these breathing modes have a chance to survive inclusion of nonlinearities. In the linearized theory, all normal mode energies $|\alpha_n|^2$ are individually conserved. Generic nonlinearities induce energy transfer between the normal modes, typically in a way that essentially involves all modes. It is unrealistic to expect that a linearized breathing mode that only depends on a subset of~$\alpha_n$ will survive in a nonlinear theory, since it is oblivious of all the other $\alpha_n$, while all the degrees of freedom participate in a complex collective dynamical process.

Breathing modes based on $b^+_{nm}$ are essentially eliminated by the above argument. Indeed, $\omega_n+\omega_m=1$ can only be satisfied for $\omega_m,\omega_n\le 1$. While it is possible to imagine artificially prepared sets of coupled oscillators with frequencies less than~1, where such breathing modes are relevant, in a realistic field theory, the normal frequencies grow without bound for short-wavelength modes. Therefore, $\omega_n\le 1$ will necessarily cover a small portion of the spectrum, and the corresponding breathing mode based on $b^+_{nm}$ will depend only on a small subset of~$\alpha_n$ and has little chance to survive in an interacting theory.

Breathing modes based on $b_{mn}$ may depend on $\alpha_n$ if there exists~$m$ such that $\omega_m=\omega_n-1$ and may depend on $\bar\alpha_n$ if there exists~$m$ such that $\omega_m=\omega_n+1$. In order for $B_0$ to depend on all $\alpha_n$ and $\bar\alpha_n$, as per the discussion above, one needs all $\omega_n$ to fit in an evenly spaced ladder
\begin{gather}
\omega_n=\omega_0+n,\label{omladder}
\end{gather}
which we shall for simplicity assume nondegenerate. All the breathing modes mentioned in the previous section are of this type. With this structure, the only nonvanishing $b_{nm}$ are $b_{n,n+1}\equiv \beta_n$, and hence we write
\begin{gather}
B_0=\sum_n \beta_n \bar\alpha_n \alpha_{n+1}.\label{B0simp}
\end{gather}
For the rest of our treatment, we shall focus on including weak nonlinearities into systems defined by (\ref{H0norm}), (\ref{omladder}) and (\ref{B0simp}).

\section{Weak nonlinearities and effective resonant dynamics}

As explained in the previous section, a natural way for a breathing mode of the form~(\ref{HBgdef}) to be supported by the evolution is to have a theory with a linearized normal mode spectrum consisting of an infinite evenly spaced ladder of the form~(\ref{omladder}). In this case, there is a linearized breathing mode of the form~(\ref{B0simp}) that one might hope to lift to the interacting theory to obtain~(\ref{HBgdef}). Assuming that has been accomplished (and our examples from Section~\ref{sec_breathe} indeed demonstrate that it is possible in special cases), what are the properties of the corresponding interacting theory in the weakly nonlinear regime $g\ll 1$?

The equations of motion arising from (\ref{HBgdef}) are
\begin{gather*}
\frac{{\rm d}\alpha_n}{{\rm d}t}={\rm i}\omega_n\alpha_n+{\rm i}g\frac{\partial H_1}{\partial\bar\alpha_n}.
\end{gather*}
It is convenient to switch to the `interaction picture' by introducing $a_n$ so that $\alpha_n=a_n {\rm e}^{{\rm i}\omega_n t}$. Then,
\begin{gather}
\frac{{\rm d}a_n}{{\rm d}t}={\rm i}g {\rm e}^{-{\rm i}\omega_nt}\frac{\partial H_1}{\partial\bar\alpha_n}\Bigg|_{\alpha_n=a_n{\rm e}^{{\rm i}\omega_n t}} = {\rm i} g\frac{\partial}{\partial \bar a_n} H_1\big(a_n{\rm e}^{{\rm i}\omega_n t},\bar a_n{\rm e}^{-{\rm i}\omega_n t}\big).\label{pstandard}
\end{gather}
The above equation is in what is known as the `periodic standard form' in mathematical literature~\cite{murdock}, which facilitates its analysis at $g\ll 1$. Qualitatively, $a_n(t)$ evolve very slowly, varying appreciably on time scales of order $1/g$, while being essentially constant on time scales of order 1. By contrast, the right-hand side contains explicit oscillatory factors ${\rm e}^{{\rm i}\omega_n t}$ varying on time scales of order~1. By the standard lore of time-averaging~\cite{murdock}, (\ref{pstandard}) can be approximated arbitrarily well for sufficiently small $g$ on long time scales of order $1/g$ by the corresponding averaged (or {\it resonant}) system of the form
\begin{gather*}
\frac{{\rm d}a_n}{{\rm d}t}={\rm i}g\frac{\partial}{\partial \bar a_n} \left(\frac1{2\pi}\int_0^{2\pi} {\rm d} t\, H_1\big(a_n{\rm e}^{{\rm i}\omega_n t},\bar a_n{\rm e}^{-{\rm i}\omega_n t}\big)\right).
\end{gather*}
Note that the $t$-integral only applies to the explicit dependence on $t$ through the oscillatory factors ${\rm e}^{\pm {\rm i}\omega_n t}$, and not to the implicit dependence on~$t$ in~$a_n$ and $\bar a_n$, which are treated as constants for the purposes of the $t$-integration. The result of the integration is an explicit function of~$a_n$ and~$\bar a_n$, while all the explicit dependence on~$t$ disappears and $g$ can be absorbed by defining the {\it slow time} $\tau=gt$. The resulting equations for $a_n$ are again in a Hamiltonian form, but with a new `resonant' Hamiltonian $H_{\rm res}$,
\begin{gather}
\frac{{\rm d}\alpha_n}{{\rm d}\tau}={\rm i}\frac{\partial H_{\rm res}}{\partial\bar\alpha_n},\qquad H_{\rm res}=\frac1{2\pi}\int\limits_0^{2\pi}{\rm d}t\, H_1\big(a_n{\rm e}^{{\rm i}\omega_n t},\bar a_n{\rm e}^{-{\rm i}\omega_n t}\big).\label{Hres}
\end{gather}
Detailed justification of the time-averaging method and the resulting resonant approximation can be found in \cite{KM, murdock}. One can more formally (and more generally) write~$H_{\rm res}$ through the evolution operator of~$H_0$ denoted as $\hat{\cal S}^t_0$, whose action on any phase space function $F$ is defined by ${\rm d}\big(\hat{\cal S}^t_0 F\big)/{\rm d}t=\big\{H_0,\hat{\cal S}^t_0F\big\}$. This gives simply
\begin{gather*}
H_{\rm res}=\frac1{2\pi}\int\limits_0^{2\pi} {\rm d}t\, \hat{\cal S}^t_0 H_1,
\end{gather*}
and we shall make use of this representation in our subsequent treatment.

We can now ask whether the breathing mode $B$ has any implications for the resonant system~(\ref{Hres}). To this end, consider the Poisson brackets $\{H_{\rm res},B_0\}$:
\begin{align*}
\{H_{\rm res},B_0\}& =\frac1{2\pi}\int\limits_0^{2\pi} {\rm d}t\, \big\{\hat{\cal S}^t_0 H_1,B_0\big\}=\frac1{2\pi}\int\limits_0^{2\pi} {\rm d}t\, \hat{\cal S}^t_0\big\{ H_1, \hat{\cal S}^{-t}_0B_0\big\}=\frac1{2\pi}\int\limits_0^{2\pi} {\rm d}t\, {\rm e}^{-{\rm i}t}\hat{\cal S}^t_0\{ H_1, B_0\} \\
& =\frac1{2\pi}\int\limits_0^{2\pi} {\rm d}t\, {\rm e}^{-{\rm i}t}\hat{\cal S}^t_0 ({\rm i}B_1-\{ H_0, B_1\} )=-\frac1{2\pi}\int\limits_0^{2\pi} {\rm d}t\, \frac{{\rm d}}{{\rm d}t} \big({\rm e}^{-{\rm i}t}\hat{\cal S}^t_0B_1\big)=\frac{B_1-\hat{\cal S}^{2\pi}_0B_1}{2\pi},
\end{align*}
where we have used the evident properties of the evolution operator $\hat{\cal S}^t\hat{\cal S}^{-t}=1$ and $\hat{\cal S}^t\{F,G\}=\big\{\hat{\cal S}^tF,\hat{\cal S}^tG\big\}$, which is easily proved by differentiating with respect to $t$ and using Jacobi identities for the Poisson brackets; in going from the first line to the second line, we used (\ref{01comm}). One can also write equivalently and more explicitly
\begin{gather}
\displaystyle\{H_{\rm res},B_0\}=\frac{B_1(a_n,\bar a_n)-B_1\big(a_n {\rm e}^{2\pi {\rm i}\omega_n},\bar a_n {\rm e}^{-2\pi {\rm i}\omega_n}\big)}{2\pi}.\label{B0Poiss}
\end{gather}
The simple expression on the right-hand side easily vanishes in special cases. For example, it would vanish if $B_1=0$, as is the case in (\ref{Bosc}), or if $\omega_0$ in (\ref{omladder}) is integer, or if $\omega_0$ is half-integer and $B_1$ is quartic in $a_n$ and $\bar a_n$. If the right-hand side of~(\ref{B0Poiss}) is zero, $B_0$ is conserved by the Hamiltonian evolution defined by $H_{\rm res}$.

The bottom line, and a key message of our treatment is then that $B_0$, the quadratic part of the breathing mode $B$ defined by (\ref{HBgdef})--(\ref{01comm}), becomes an ordinary {\it conserved quantity} within the resonant approximation at~$g\ll 1$, provided that
\begin{gather}
B_1(a_n,\bar a_n)=B_1\big(a_n {\rm e}^{2\pi {\rm i}\omega_0},\bar a_n {\rm e}^{-2\pi {\rm i}\omega_0}\big).
\label{condB1}
\end{gather}
This statement is independent of the form of $B_1$ and $H_1$, as long as they satisfy the definition~(\ref{01comm}), and the linearized breathing mode $B_0$ is realized as (\ref{omladder})--(\ref{B0simp}). The specific examples of breathing modes given in section \ref{sec_breathe} follow this pattern (at least after the evolution has been truncated to appropriate subsets of modes labelled by a single integer~$n$).

Assume now that (\ref{condB1}) is satisfied so that $B_0$ of the form (\ref{B0simp}) is a conserved quantity of~$H_{\rm res}$. If~$H_1$, and hence $H_{\rm res}$, is quartic in $a_n$ and $\bar a_n$, further implications of the breathing mode can be exposed. The most general quartic~$H_{\rm res}$ one could write is
\begin{gather}
H_{\rm res}= \sum_{\omega_n+\omega_m=\omega_k+\omega_l} C_{nmkl}\bar a_n \bar a_m a_k a_l + \sum_{\omega_n=\omega_m+\omega_k+\omega_l} S_{nmkl}\bar a_n a_m a_k a_l \nonumber\\
\hphantom{H_{\rm res}=}{} + \sum_{\omega_n=\omega_m+\omega_k+\omega_l} \bar S_{nmkl} a_n \bar a_m \bar a_k \bar a_l,\label{Hresgeneral}
\end{gather}
where $C$ and $S$ are numerical coefficients. Terms with four $a$'s or four $\bar a$'s could not possibly survive the time averaging in the definition~(\ref{Hres}). Assuming that~(\ref{condB1}) is satisfied and hence~$B_0$ of the form~(\ref{B0simp}) is a conserved quantity of $H_{\rm res}$, the following symmetry transformation associated to $B_0$ must be respected by $H_{\rm res}$:
\begin{gather}
a_n\to a_n+ {\rm i}\eta \beta_n a_{n+1} + {\rm i}\bar\eta\bar\beta_{n-1} a_{n-1},\label{atrans}
\end{gather}
which imposes relations between the coefficients $C_{nmkl}$ and $S_{nmkl}$ in $H_{\rm res}$. (In all of our formulas it should be understood that if a mode number index is outside the standard range $[0,\infty)$, the corresponding expression is~0.) The actual form of the constraints on $C$ and $S$ from the above symmetry transformations is
\begin{gather}
\beta_n C_{n+1,m,k,l}+\beta_m C_{n,m+1,k,l}=\beta_{k-1} C_{n,m,k-1,l}+ \beta_{l-1} C_{n,m,k,l-1},\label{Cbeta}\\
\beta_n S_{n+1,m,k,l}=\beta_{m-1} S_{n,m-1,k,l}+\beta_{k-1} S_{n,m,k-1,l}+ \beta_{l-1} S_{n,m,k,l-1}.\label{Sbeta}
\end{gather}
The equation for $S$, in fact, guarantees that $S_{nmkl}=0$. Indeed, setting $m=k=l=0$ in~(\ref{Sbeta}), we obtain $S_{n000}=0$, and then one proceeds recursively increasing $m$, $k$ and $l$ in steps of~1 to prove that $S_{nmkl}=0$. This is closely related to the selection rules for AdS mode couplings discussed in~\cite{EN}. With only $C$ in place, the resonant Hamiltonian takes the simple form
\begin{gather}
H_{\rm res}= \sum_{n+m=k+l} C_{nmkl} \bar a_n \bar a_m a_k a_l,\label{Hressimp}
\end{gather}
which is familiar from \cite{BBCE,BBCE1,AO,AO1,CF,BEF, BEL}. Note that, with $S$ having dropped out, the resonant Hamiltonian enjoys two conservation laws
\begin{gather}
N=\sum_n |a_n|^2,\qquad E=\sum_n n|a_n|^2,\label{NE}
\end{gather}
irrespectively of the values of $C$.

We have just seen that, for a system with quartic nonlinearities, if~(\ref{condB1}) is satisfied and $B_0$ becomes a conserved quantity of $H_{\rm res}$, a number of possible terms in $H_{\rm res}$ drop out, leaving the simple expression~(\ref{Hressimp}). The converse is also true: if it happens that
the $S$-couplings in~(\ref{Hresgeneral}) vanish for a specific quartic system, and the resonant Hamiltonian is of the form~(\ref{Hressimp}), then~(\ref{condB1}) is satisfied, and hence $B_0$ becomes a conserved quantity of $H_{\rm res}$. Indeed, $H_{\rm res}$ of~(\ref{Hressimp}) is bilinear in~$a$ and bilinear in~$\bar a$, while $B_0$ is linear in~$a$ and linear in $\bar a$. Therefore, $\{H_{\rm res},B_0\}$ is likewise bilinear in $a$ and bilinear in $\bar a$, and should arise, by~(\ref{B0Poiss}), from terms in~$B_1$ bilinear in $a$ and bilinear in $\bar a$. But any such terms would give a vanishing contribution to $B_1(a_n,\bar a_n)-B_1\big(a_n {\rm e}^{2\pi {\rm i}\omega_0},\bar a_n {\rm e}^{-2\pi {\rm i}\omega_0}\big)$, leaving nothing on the right-hand side of~(\ref{B0Poiss}), and yielding $\{H_{\rm res},B_0\}=0$.

We now reexamine the breathing mode (\ref{B0simp}) that has become a conserved quantity of (\ref{Hressimp}). If $B_0$ is a conserved quantity of $H_{\rm res}$, so are $\bar B_0$ and $\big\{\bar B_0,B_0\big\}$, which is explicitly given by
\begin{gather*}
\big\{\bar B_0,B_0\big\}={\rm i}\sum_{n=0}^{\infty}\big(|\beta_n|^2-|\beta_{n-1}|^2\big)|a_n|^2.
\end{gather*}
This conserved quantity is itself of a form similar to~(\ref{NE}), being a weighted sum of the individual linearized mode energies~$|a_n|^2$. Each such conserved quantity constrains the way nonlinearities may dynamically redistribute the energy among the normal modes. It may be reasonable to demand that no further constraints of this sort, beyond the generic conservation of $N$ and $E$, are present. In this case, $\big\{\bar B_0,B_0\big\}$ must be a linear combination of $N$ and $E$, which we can write as
\begin{gather*}
\big\{\bar B_0,B_0\big\}={\rm i}\left(N+\frac{2E}{G}\right).
\end{gather*}
Here, $G$ is an arbitrary number, while the numerical coefficient in front of $N$ has been set to~1 as a matter of fixing the normalization of~$B_0$, which has been until now kept undetermined. One then has
\begin{gather*}
|\beta_n|^2-|\beta_{n-1}|^2=1+\frac{2n}G,
\end{gather*}
which is solved by
\begin{gather*}
|\beta_n|^2=(1+n)(1+n/G).
\end{gather*}
As the phases of $\beta_n$ can be arbitrarily shifted by adjusting the phases of $a_n$, one can simply define $\beta_n$ to be the square root of the right-hand side,
\begin{gather*}
\beta_n=\sqrt{(1+n)(1+n/G)},
\end{gather*}
reducing $B_0$ to
\begin{gather}
B_0=\sum_n \sqrt{(1+n)(1+n/G)} \bar a_n a_{n+1}.\label{B0AO}
\end{gather}
Thus, with a series of simple and generic assumptions on how the breathing mode is realized in the linearized theory, how simple conditions are met to promote the breathing mode to a conserved quantity of the resonant approximation to the weakly nonlinear theory, and how taking Poisson brackets of the breathing mode with its own complex conjugate does not generate new conserved quantities, we have arrived at the class of `solvable' resonant systems developed in~\cite{AO,AO1}. Indeed, the resonant Hamiltonian~(\ref{Hressimp}) explicitly matches the constructions of~\cite{AO,AO1}, while the conserved quantity $B_0$ of~(\ref{B0AO}) corresponds, in the notation of \cite{AO,AO1} to $\bar Z/\sqrt{G}$. We shall therefore conclude by simply restating the consequences of (\ref{Hressimp}) and~(\ref{B0AO}) already explored in \cite{AO,AO1}.

With $\beta_n=\sqrt{(1+n)(1+n/G)}$, (\ref{Cbeta}) imposes constraints on the coefficients of the resonant system (\ref{Hressimp}), which are identical to the ones used in \cite{AO,AO1} to define the `solvable' class of resonant system. `Solvability' is understood here in a very restricted sense, namely, as having an explicit family of nontrivial solutions. This family is defined by the ansatz
\begin{gather}
a_n(t)= \sqrt{\frac{G(G+1)(G+2)\cdots (G+n-1)}{G^n n!}}(b(t)+a(t)n)(p(t))^n,\label{ansatz}
\end{gather}
where $b(t)$, $a(t)$ and $p(t)$ are complex-valued functions of time (and the conventions in the above formula differ slightly and inessentially from \cite{AO,AO1}). The Hamiltonian equations of motion of~(\ref{Hressimp}) are
\begin{gather}
\frac{da_n}{dt}={\rm i}\sum_{m=0}^\infty\sum_{k=0}^{n+m}C_{nmk,n+m-k} \bar a_m a_k a_{n+m-k}.\label{eoma}
\end{gather}
It is a nontrivial fact that the ansatz (\ref{ansatz}) is consistent with these equation of motion, and yet it is true by virtue of the conservation of~$B_0$ and the identity~(\ref{Cbeta}) it implies, as demonstrated in~\cite{AO,AO1}. A key point of the proof is that finite-difference identities~(\ref{Cbeta}) imply summation identities for $C_{nmkl}$ adapted to the summation structure in~(\ref{eoma}). As a result, one obtains a~closed system of three ODEs for $b(t)$, $a(t)$ and $p(t)$, which is furthermore superintegrable because of the conservation of $H$, $N$, $E$ and $B_0$. The ODEs can be integrated to show that $|p(t)|$ is always a strictly periodic function for all solutions, and the same is true for the spectrum~$|a_n|^2$. An explicit bound can be given on the turbulent transfer of energy toward large $n$ modes for the solutions in the ansatz~(\ref{ansatz}). The reader is referred to~\cite{AO,AO1} for detailed derivations of these results.

It is worth noting that the above properties were developed in \cite{AO,AO1} completely in the language of resonant systems of the form (\ref{Hressimp}), without any specific attention to how such features could emerge in resonant systems arising as weakly nonlinear approximations to realistic PDEs. Our present treatment closes this gap. We also remark that it is outside the normal range of implications of symmetries that explicit families of solutions, as given by (\ref{ansatz}), are generated. Symmetries produce new solutions out of known solutions, but (\ref{ansatz}) does not follow by application of transformations (\ref{atrans}) to any other, more obvious solutions of (\ref{eoma}). Rather, the logic here is that the identities (\ref{Cbeta}) imposed on the mode couplings by the symmetries have further implications and allow for the closure of the ansatz (\ref{ansatz}). This feature is specific to quartic nonlinearities, and does not immediately generalize to other cases.

\section{Discussion}

We have revisited the topic of breathing modes in the dynamics of nonlinear PDEs, and in particular, the implications of the breathing modes for the weakly nonlinear regime. We have assumed that both the Hamiltonian and the breathing mode are linear functions of a coupling parameter, and that setting the coupling parameter to zero results in a linear dynamical system, with a~quadratic Hamiltonian, wherein the breathing mode also becomes quadratic in the canonical variables, which corresponds to a linear realization of the corresponding kinematic symmetry. Such setup is very generic from a physical perspective, commonly occurring in classical field theories. We have presented a~collection of explicit breathing modes related to the dynamics of Bose--Einstein condensates and anti-de Sitter spacetimes. While the breathing modes (\ref{Bosc}) and (\ref{BPR}) are standard in the Bose--Einstein literature, the corresponding relativistic breathing modes (\ref{breathAdS}) and (\ref{BCF}) are in principle known from the symmetry properties of AdS spacetimes, but we believe our explicit expressions are compact and convenient.

We have discussed how breathing modes of our type may be realized in a linear theory. The most natural realization is for systems whose normal mode frequencies form evenly spaced ladders, as in (\ref{omladder}). Such an evenly-spaced spectrum is highly resonant and, by the standard lore of weakly nonlinear dynamics, creates a possibility for arbitrarily small nonlinearities of order $g$ to produce arbitrarily large effects on time scales of order $1/g$. On these specific time-scales, the original dynamics may be accurately approximated by the time-averaged dynamics, described by the resonant system (\ref{Hres}). A simple condition (\ref{condB1}), which is easily satisfied in special cases of interest, ensures that the quadratic part of the original breathing mode becomes a conserved quantity of the effective resonant dynamics (\ref{Hres}).

If the nonlinearities are quartic, as is common in field theories, further consequences result from the conservation law in the resonant system inherited from the breathing mode of the original system. First, only one of the possible quartic terms may remain in the resonant Hamiltonian, leaving a simple expression (\ref{Hressimp}). Two conservation laws (\ref{NE}) are then obeyed by the resonant system. Assuming that the algebra of conserved quantities closes on the resonant Hamiltonian, the breathing mode and these two extra quantities fixes the functional form of the breathing mode in terms of one free parameter (\ref{B0AO}). This recovers, starting from physically motivated PDE problems, resonant systems of the solvable class considered in \cite{AO,AO1}. As a consequence, one obtains explicit solutions of the form (\ref{ansatz}) at the level of the resonant approximation, which can be thoroughly analyzed as in \cite{AO,AO1}.

Our treatment explains in a uniform fashion the emergence of solvable features within the resonant approximation in a number of physically motivated PDEs in the recent literature \cite{BBCE,BBCE1,CF,BEF,BEL,GGT}. In particular, the progenitors of these solvable features in the resonant systems are identified as breathing modes in the PDEs whose dynamics the resonant systems approximate. With respect to the solvable resonant systems of \cite{AO,AO1}, our treatment provides a mechanism by which they can emerge as approximations to specific PDEs of mathematical physics. Systems with breathing modes may be engineered starting with linear systems whose normal frequencies form evenly spaced ladders (\ref{omladder}), which creates a lot of room for concrete applications of our analysis. In relation to the concrete physical problems that have motivated our considerations, beyond what has been explicitly treated in the literature, one is led to expect solvable features in the resonant systems corresponding to (1) one-dimensional nonlinear Schr\"odinger equation in a harmonic trap with arbitrary 2-body interactions, (2) Landau-level truncations, in the style of \cite{BBCE,BBCE1}, of nonlinear Schr\"odinger equations in isotropic harmonic traps with arbitrary 2-body interactions in any number of dimensions, (3) maximally rotating truncations of the resonant dynamics in AdS, in the style of \cite{BEL}, with arbitrary quartic local interactions. The last topic connects to extensive studies of nonlinear dynamics in AdS \cite{FPU,BMR,BR,squash,rev2,CEV,CEV1}, in particular, outside spherical symmetry \cite{ads4,ads3,ads5,ads2,ads1}. Some of the results presented here, in particular explicit analytic solutions within the resonant approximation, are specific to the case of quartic nonlinearities. It would be interesting to investigate whether generalizations of these results (which are expected to be non-straightforward) exist for more general nonlinearities.

 {\em Note added:} An anonymous referee has aptly observed that the last condition listed in~(\ref{01comm}), namely $\{H_1,B_1\}=0$, has never been used in our derivations. This means that, technically, it is sufficient for the breathing mode definition (\ref{breathdef}) to be satisfied up to linear order in $g$ to ensure that the formalism developed here is applicable.

\subsection*{Acknowledgments}

I have benefitted from discussions with Anxo Biasi, Piotr Bizo\'n, Ben Craps and Andrzej Rostworowski.
This research is supported by CUniverse research promotion project at Chulalongkorn University (grant CUAASC) and by FWO-Vlaanderen through project G006918N. Part of this work was developed during a visit to the physics department of the Jagiellonian University (Krakow, Poland). Support of the Polish National Science Centre through grant number 2017/26/A/ST2/00530 and personal hospitality of Piotr and Magda Bizo\'n are gratefully acknowledged.

\pdfbookmark[1]{References}{ref}
\LastPageEnding

\end{document}